\documentstyle[preprint,aps]{revtex}
\tighten
\draft
\widetext
\preprint{YITP-00-35}
\bigskip
\bigskip
\begin{document}
\title{Conformal Brane World and Cosmological Constant}
\medskip
\author{Zurab Kakushadze\footnote{E-mail: 
zurab@insti.physics.sunysb.edu}}
\bigskip
\address{C.N. Yang Institute for Theoretical Physics\\ 
State University of New York, Stony Brook, NY 11794}

\date{August 3, 2000}
\bigskip
\medskip
\maketitle

\begin{abstract} 
{}We consider a recently proposed setup where a codimension one brane is
embedded in the background of a smooth domain wall interpolating between
AdS and Minkowski minima. Since the volume of the transverse dimension is
infinite, bulk supersymmetry is intact even if brane supersymmetry is
completely broken. On the other hand, in this setup unbroken bulk 
supersymmetry is incompatible with non-zero brane cosmological constant,
so the former appears to protect the latter. In this paper we point out that,
to have a consistent coupling between matter localized on the brane and bulk
gravity, in this setup generically
it appears to be necessary that the brane world-volume
theory be conformal. Thus, unbroken bulk supersymmetry appears to actually
protect not only the cosmological constant but also 
conformal invariance on the brane.  
\end{abstract}

\section{Introduction}

{}In the Brane World scenario the Standard Model gauge and matter fields
are assumed to be localized on  
branes (or an intersection thereof), while gravity lives in a larger
dimensional bulk of space-time 
\cite{early,BK,polchi,witt,lyk,shif,TeV,dienes,3gen,anto,ST,BW}. The volume
of dimensions transverse to the branes is 
automatically finite if these dimensions are compact. On the other hand, 
the volume of the transverse dimensions
can be finite even if the latter are non-compact. In particular, this can be
achieved by using \cite{Gog} warped compactifications \cite{Visser} which
localize gravity on the brane. A concrete
realization of this idea was given in \cite{RS}.

{}Recently it was pointed out in
\cite{DGP0,witten} that, in theories where extra dimensions
transverse to a brane have infinite volume\cite{GRS,CEH,DVALI,DGP1,DGP}, 
the cosmological constant on the
brane might be under control even if brane supersymmetry is completely
broken. The key point here is that even if supersymmetry breaking on the
brane does take place, it will not be transmitted to the bulk as the volume 
of the extra dimensions is infinite \cite{DGP0,witten}. Thus, at least in 
principle, one should be able to control some of the properties of the bulk
with the unbroken bulk supersymmetry. One then can wonder whether bulk 
supersymmetry could also control the brane cosmological constant
\cite{DGP0,witten}.

{}Controlling the brane cosmological constant with bulk supersymmetry,
however, appears to be non-trivial. Thus, in \cite{zura} it was pointed out 
that in the Dvali-Gabadadze-Porrati model \cite{DGP}, where a 3-brane is
embedded in the 5-dimensional Minkowski space, unbroken bulk supersymmetry
is perfectly compatible with non-vanishing, in particular, positive brane
cosmological constant. There is a simple reason for this. The bulk
curvature in this model is constant, in particular, it vanishes. The Minkowski 
space can be foliated by codimension one surfaces of vanishing or positive
constant curvature. Both of these types of foliations are compatible with bulk
supersymmetry, that is, with the existence of Killing spinors in the bulk -
the latter is a (local) property of the corresponding space-time, and is 
independent of the foliation. Note that this also applies to another space
of constant curvature which admits Killing spinors, namely, the AdS space.
In this case we have foliations with vanishing, positive and negative
constant curvature, all of which are perfectly consistent with bulk 
supersymmetry. 

{}It is then natural to consider examples where the bulk is a space with
non-constant curvature, which, nonetheless, admits Killing spinors. It is
clear that (in the codimension one case) such a space would have half of
the supersymmetries compared with the constant curvature cases. One way to
parametrize such a space is to consider a bulk theory where gravity is
coupled to a scalar field $\phi$ with a non-trivial scalar potential
$V(\phi)$ (that is, $V(\phi)$ is not a constant). For an appropriately
chosen scalar potential there exist BPS domain wall solutions which
preserve half of the original supersymmetries (that is, half of the 
supersymmetries corresponding to Minkowski or AdS minima of the
scalar potential). The corresponding foliation is then necessarily 
flat. The reason why this is so different from the cases where the bulk 
curvature is constant is that in the latter cases one only needs to ensure
vanishing of the gravitino variation, while in the presence of a non-trivial
bulk potential preserving bulk supersymmetry also requires vanishing of
the variation of the superpartner of $\phi$.

{}This is precisely the proposal of \cite{zura1}. More concretely, let a
codimension one brane (which, for simplicity, is taken to be 
$\delta$-function-like) be embedded in the BPS domain wall background of
the aforementioned type. To have an infinite volume extra dimension, the
scalar potential is assumed to have one Minkowski minimum and one AdS
minimum, and the domain wall interpolates between these minima. Note that
such a domain wall (unlike a domain wall interpolating between two
Minkowski minima \cite{witten,CEH}) does not have to violate the weak energy 
condition. In particular, if the domain wall is smooth, which is the
case if the brane cosmological constant equals the brane tension, then the
weak energy condition is not violated \cite{zura1}. Now, since the volume
of the extra dimension is infinite, supersymmetry breaking on the brane 
does not result in bulk supersymmetry breaking. However, the latter is 
not compatible with non-zero brane cosmological constant. Thus, even if 
supersymmetry breaking on the brane does occur, the brane cosmological 
constant appears to remain zero.

{}The purpose of this paper, which is essentially a follow-up of
\cite{zura1}, is to point out that the theory living on the
brane in the setup of \cite{zura1} generically
appears to be conformal. One way to see
this is to consider small fluctuations around the background in the presence
of matter localized on the brane. Then the system of equations for the small
fluctuations of the scalar field and the metric is essentially 
overconstrained, and has consistent solutions if and only if the 
energy-momentum tensor of the matter localized on the brane is traceless
and the coupling of the scalar field to the brane matter is vanishing.
This appears to be due to the fact that a tensionless brane by itself does not
explicitly break diffeomorphism invariance, so that the latter is broken 
spontaneously by the domain wall solution. Spontaneous breaking of
translational invariance results in the gravitational Higgs mechanism
discussed in \cite{COSM}, in the process of which the trace part 
$h^\mu_\mu$ of the graviton roughly 
becomes a pure gauge. This then implies that
in such a setup a consistent coupling of bulk gravity to the brane matter is
possible only if the latter is conformal.

{}Thus, the approach of \cite{zura1} might provide a realization
of a setup where a conformal theory on the brane is gravitationally 
coupled to a non-conformal theory in the bulk. The conformal property
of the brane world-volume theory is then preserved due to bulk supersymmetry,
which is unbroken even if the brane theory is non-supersymmetric as the volume
of the extra dimension is infinite. In particular, if gravitational
interactions localized on the brane are generated via, say, loop diagrams,
then the corresponding brane world gravity is expected to be conformal
as well.
 
{}The remainder of this paper is organized as follows. In section II
we briefly review the setup of \cite{zura1}. In section III we study 
small fluctuations around the solution in the presence of brane matter
sources, and discuss the requirement that the brane
matter be conformal. Section IV contains concluding remarks.
 
\section{The Setup}

{}In this section we review the setup of \cite{zura1}. Thus, consider the
model with the following action (more precisely, here we give the part of
the action relevant for the subsequent discussions):
\begin{equation}\label{action}
 S={\widehat M}_P^{D-3}\int_\Sigma d^{D-1} x \sqrt{-{\widehat G}}\left[
 {\widehat R}-{\widehat\Lambda}\right] +
 M_P^{D-2}
 \int d^D x \sqrt{-G} \left[R-{4\over{D-2}}(\nabla\phi)^2-V(\phi) \right]~.
\end{equation}
For calculational convenience we will keep the number of space-time
dimensions $D$ unspecified.
In (\ref{action}) ${\widehat M}_P$ is (up to a normalization factor - see 
below) the $(D-1)$-dimensional (reduced) Planck scale, while $M_P$ is the 
$D$-dimensional one. The $(D-1)$-dimensional hypersurface $\Sigma$, which we
will refer to as the brane, is the $y=y_0$ slice of the $D$-dimensional 
space-time,
where $y\equiv x^D$, and $y_0$ is a constant. Next, 
\begin{equation}
 {\widehat G}_{\mu\nu}\equiv{\delta_\mu}^M {\delta_\nu}^N G_{MN}
 \Big|_{y=y_0}~,
\end{equation} 
where the capital Latin indices $M,N,\dots=1,\dots,D$, while the Greek
indices $\mu,\nu,\dots=1,\dots,(D-1)$. The quantity ${\widehat\Lambda}$ is
the brane tension. More precisely, there might be various (massless and/or 
massive) fields (such
as scalars, fermions, gauge vector bosons, {\em etc.}), which we
will collectively denote via $\Phi^i$, localized on the brane. Then ${\widehat
\Lambda}={\widehat\Lambda}(\Phi^i,\nabla_\mu\Phi^i,\dots)$ generally depends
on the vacuum expectation values of these fields as well as their derivatives.
In the following we will assume that the expectation values of the $\Phi^i$
fields are dynamically determined, independent of the coordinates 
$x^\mu$, and consistent with $(D-1)$-dimensional general covariance. 
The quantity ${\widehat\Lambda}$ is then a constant which we identify 
as the brane tension. 
The bulk fields are given by the metric $G_{MN}$, a single
real scalar field $\phi$, as well as other fields (whose expectation values
we assume to be vanishing) which would appear in a concrete supergravity 
model (for the standard values of $D$). 

{}Let us briefly comment on the 
$\sqrt{-{\widehat G}} {\widehat R}$ term in the brane world-volume action.
Typically such a term is not included in discussions of various brane world
scenarios (albeit usually the $-\sqrt{-{\widehat G}} {\widehat \Lambda}$ 
term is). 
However, as was pointed out in \cite{DGP}, even if such a term is absent at the
tree level, as long as the 
brane world-volume theory is not conformal, it will typically be
generated by quantum loops of other fields localized on the 
brane (albeit 
not necessarily with the desired sign, which, nonetheless, appears to be as
generic as the opposite one). This is an important observation, which allows 
to reproduce the $(D-1)$-dimensional Newton's law on, say,
a non-conformal brane embedded in $D$-dimensional Minkowski space-time 
\cite{DGP}. However, as we will see in the following, in the above setup the
brane world-volume theory is actually conformal, and ${\widehat M}_P=0$.

{}To proceed further, we will need equations of motion following from the
action (\ref{action}). Here we are interested in studying possible solutions
to these equations which are consistent with $(D-1)$-dimensional general 
covariance. That is, we will be looking for solutions with the warped
metric of the following form:
\begin{equation}\label{warped}
 ds_D^2=\exp(2A){\widetilde g}_{\mu\nu}dx^\mu dx^\nu +dy^2~,
\end{equation}
where the warp factor $A$ and the scalar field $\phi$, 
which are functions of $y$,
are independent of the coordinates
$x^\mu$, and the $(D-1)$-dimensional metric 
${\widetilde g}_{\mu\nu}$ is independent of $y$. With this 
ans{\"a}tz, we have the following
equations of motion for $\phi$ and $A$:
\begin{eqnarray}\label{phi''d}
 &&{8\over {D-2}}\left[\phi^{\prime\prime}+(D-1)A^\prime\phi^\prime\right]-
 V_\phi-L f_\phi\delta(y-y_0)=0~,\\
 \label{phi'A'd}
 &&(D-1)(D-2)(A^\prime)^2-{4\over{D-2}}(\phi^\prime)^2+V-
 {{D-1}\over{D-3}}{\widetilde \Lambda}\exp(-2A)=0~,\\
 \label{A''d}
 &&(D-2)A^{\prime\prime}+{4\over {D-2}}(\phi^\prime)^2+{1\over {D-3}}
 {\widetilde \Lambda}\exp(-2A)+{1\over 2} L f\delta(y-y_0)=0~.
\end{eqnarray}
Here 
\begin{equation}
 f\equiv{\widehat \Lambda}- {\widetilde \Lambda}\exp[-2A(y_0)]
\end{equation}
is the effective brane tension.
The scale $L$, defined as
\begin{equation}
 L\equiv {\widehat M}_P^{D-3}/M_P^{D-2}~,
\end{equation}
plays the role of the crossover distance scale below which gravity is
effectively $(D-1)$-dimensional, while above this scale it becomes 
$D$-dimensional. Next, ${\widetilde \Lambda}$ 
is independent of $x^\mu$ and $y$. In fact, it 
is nothing but the cosmological constant of the $(D-1)$-dimensional manifold,
which is therefore an Einstein manifold, corresponding to the hypersurface
$\Sigma$. Our normalization of ${\widetilde\Lambda}$ is such that
the $(D-1)$-dimensional metric ${\widetilde g}_{\mu\nu}$ satisfies
Einstein's equations:
\begin{equation}
 {\widetilde R}_{\mu\nu}-{1\over 2}{\widetilde g}_{\mu\nu}
 {\widetilde R}=-{1\over 2}
{\widetilde g}_{\mu\nu}{\widetilde\Lambda}~.
\end{equation}

{}Here we note that in the bulk (that is, for $y\not=y_0$) one of the
second order equations is automatically satisfied once the first
order equation (\ref{phi'A'd}) as well as the other second order equation are
satisfied. As usual, this is a consequence of Bianchi identities.

{}Note that by rescaling the coordinates $x^\mu$ on the brane we can always
set $\exp[A(y_0)]=1$. Then the $(D-1)$-dimensional Planck scale is simply
${\widehat M}_P$. Let $\phi_0\equiv\phi(y_0)$.
Note that the above system of equations has smooth solutions for 
\begin{equation}
 f(\phi_0)=f_\phi(\phi_0)=0~,
\end{equation}
that is, if the brane cosmological constant and the brane tension are equal
\begin{equation}
 {\widetilde\Lambda}={\widehat\Lambda}~,
\end{equation} 
and there is no $\phi$ tadpole due to the brane.
In particular, in these solutions $\phi$ and $A$ as well as their derivatives 
$\phi^\prime$ and $A^\prime$ are smooth. 

{}Let us now discuss possible solutions of the above system of equations 
(\ref{phi''d}), (\ref{phi'A'd}) and ({\ref{A''d}) for
$f(\phi_0)=f_\phi(\phi_0)=0$. 
To obtain an infinite volume solution, let us assume that the scalar
potential has one AdS minimum located at $\phi=\phi_-$ and one Minkowski
minimum located at $\phi=\phi_+$ (without loss of generality we will assume
that $\phi_+>\phi_-$). Moreover, let us assume that there are no other 
extrema except for a dS maximum located at $\phi=\phi_*$, where $\phi_-<\phi_*
<\phi_+$, such that $V(\phi_*)\gg V(\phi_+)-V(\phi_-)=|V(\phi_-)|$. This latter
condition is necessary to sufficiently suppress the probability for nucleation
of AdS bubbles in the Minkowski vacuum, which could otherwise destabilize
the background \cite{okun}. Then 
we can have smooth domain walls interpolating between the two vacua. In fact,
for ${\widetilde\Lambda}=0$ we have $\phi(y)\rightarrow\phi_\pm$ as 
$y\rightarrow\pm\infty$. On the other hand, for ${\widetilde\Lambda}>0$ we have
$\phi(y)\rightarrow\phi_+$ as $y\rightarrow+\infty$, while 
$\phi(y)\rightarrow\phi_-$ as $y\rightarrow y_-$, where $y_-<y_0$ is finite 
(here for definiteness we have assumed that the domain wall approaches the 
Minkowski vacuum as $y\rightarrow+\infty$). As to the warp factor $A$,
it goes to $-\infty$ as $\phi\rightarrow
\phi_-$ (if ${\widetilde
\Lambda}=0$, then $A$ goes to $-\infty$ linearly with $|y|$, while if 
${\widetilde \Lambda}>0$, then $A\sim\ln(y-y_-)$ as $y\rightarrow y_-$). On 
the other hand, if ${\widetilde \Lambda}=0$, then $A$ goes to a constant
as $y\rightarrow+\infty$, while if ${\widetilde\Lambda}>0$, then $A$ grows
logarithmically with $y$. In both cases the volume of the extra dimension is
infinite as the integral 
\begin{equation}
 \int dy \exp[(D-1)A]
\end{equation}
diverges. Moreover, there are no quadratically normalizable bulk graviton 
modes. Rather, for ${\widetilde\Lambda}=0$ we have a continuum of plane-wave
normalizable bulk modes (with mass squared $m^2\geq 0$), while for
${\widetilde\Lambda}>0$ we have a mass gap in the bulk graviton spectrum,
and the plane-wave normalizable modes are those with $m^2>m_1^2$, where
$m_1^2\sim{\widetilde\Lambda}$ \cite{zura}. Thus, without any additional 
assumptions consistent solutions with vanishing as well as positive 
brane cosmological constant exist for such potentials.
   
{}However, as was pointed out in \cite{zura1}, 
as long as the scalar potential $V(\phi)$ is
non-trivial, bulk supersymmetry is incompatible with non-zero brane
cosmological constant. Indeed, this immediately follows from the bulk Killing
spinor equations (following from the requirement that variations of the
superpartner $\lambda$ of $\phi$ and the gravitino $\psi_M$ vanish
under the corresponding supersymmetry transformations), which in such 
backgrounds reduce to:
\begin{eqnarray}\label{killing1}
 &&\phi^\prime=\alpha W_\phi~,\\
 \label{killing2}
 &&A^\prime=\beta W~,
\end{eqnarray}
where $W$ is the superpotential,
\begin{equation}
 \alpha\equiv\eta{\sqrt{D-2}\over 2}~,~~~\beta\equiv-\eta
 {2\over (D-2)^{3/2}}~,
\end{equation}
and $\eta=\pm 1$.

{}Note that the system of equations (\ref{killing1}) and (\ref{killing2})
is compatible with the system of equations (\ref{phi''d}), 
(\ref{phi'A'd}) and (\ref{A''d}) if and only if ${\widetilde\Lambda}=0$,
and the scalar potential is given by
\begin{equation}
 V=W_\phi^2-\gamma^2 W^2~,
\end{equation}
where 
\begin{equation}
 \gamma\equiv {2\sqrt{D-1}\over{D-2}}~.
\end{equation}
Thus, bulk supersymmetry (note that the domain wall solution preserves
$1/2$ of the supersymmetries corresponding to the minima of $V$) is preserved
if and only if the brane cosmological constant vanishes. We therefore
conclude that even if brane supersymmetry is broken, bulk supersymmetry,
which remains unbroken as the volume of the transverse dimension is
infinite, ensures that the brane cosmological constant still vanishes
in the model defined in (\ref{action}).

{}Before we end this section, for illustrative purposes let us give an
example of a domain wall of the aforementioned type. Let
\begin{equation}\label{sup1}
 W=\xi\left[v^2\phi-{1\over 3}\phi^3-{2\over 3} v^3\right]~.
\end{equation}
Note that at $\phi_-=-v$ we have 
the AdS minimum, while at $\phi_+=+v$ we have the Minkowski minimum. To ensure
that the condition $|V(\phi_-)|\ll
V(\phi_*)$ is satisfied, where $\phi_*$ ($\phi_-<\phi_*<\phi_+$) 
corresponds to the dS maximum, we must assume 
$v\ll 1$. The domain wall solution, which interpolates between the AdS and
Minkowski vacua in
this case is given by
\begin{eqnarray}
 &&\phi(y)=v\tanh(\alpha\xi v(y-y_1))~,\\
 &&A(y)={2\beta\over3\alpha}v^2 \left[\ln(\cosh(\alpha\xi v(y-y_1)))-
 {1\over 4}{1\over{\cosh^2(\alpha\xi v(y-y_1))}}\right]-
 {2\beta\over 3}\xi v^3(y-y_1)+C~,
\end{eqnarray}  
where $y_1$ and $C$ are integration constants.

{}Finally, let us note that solutions with non-vanishing $f(\phi_0)$ and
$f_\phi(\phi_0)$ do not interpolate between the AdS and Minkowski vacua.
Thus, solutions with positive $f(\phi_0)$ asymptotically approach the AdS
minimum on both sides of the brane, while solutions with negative  
$f(\phi_0)$ asymptotically approach the Minkowski
minimum on both sides of the brane\footnote{Note that $f(\phi_0)$ is the
effective brane tension. If $f(\phi_0)<0$, then we have world-volume 
ghosts unless we
assume that the brane is an ``end-of-the-world'' brane located at an orbifold
fixed point. Thus, in solutions with $f(\phi_0)<0$ 
the geometry of the $y$ dimension
is that of ${\bf R}/{\bf Z}_2$ (and not of ${\bf R}$), with the orbifold
fixed point identified with $y_0$ (then the corresponding solution 
on the covering space has the 
${\bf Z}_2$ symmetry required for the orbifold interpretation), and the
brane is stuck at the orbifold fixed point.}.

\section{Brane Matter Sources}

{}In this section we would like to study
gravitational interactions between sources localized on the brane. To do
this, let us start from the action (\ref{action}), and study small 
fluctuations of the metric $G_{MN}$ and the scalar field $\phi$, 
which we will denote via $h_{MN}$ and $\varphi$, respectively, around the
corresponding smooth domain wall solution (with vanishing brane cosmological 
constant) in the presence of brane matter sources. 

{}In the following it will prove convenient to make the coordinate 
transformation $y\rightarrow z$ so that the background metric takes
the form:
\begin{equation}
 ds_D^2=\exp(2A)\left[\eta_{\mu\nu}dx^\mu dx^\nu+dz^2\right]~.
\end{equation}
That is,
\begin{equation}\label{yz}
 dy=\exp(A)dz~,
\end{equation}
where we have chosen the overall sign so that $z\rightarrow\pm\infty$ as
$y\rightarrow\pm\infty$. Moreover, we can fix the integration constant 
upon solving (\ref{yz}) such that $y=y_0$ is mapped to $z=0$.
So from now on we will use the coordinates $x^M=(x^\mu,x^D)=(x^\mu,z)$,
and prime will denote derivative w.r.t. $z$. Moreover, the capital
Latin indices $M,N,\dots$ are lowered and raised with the flat $D$-dimensional
Minkowski metric $\eta_{MN}$ and its inverse, while the Greek indices
$\mu,\nu,\dots$ are lowered and raised with the flat $(D-1)$-dimensional 
Minkowski metric $\eta_{\mu\nu}$ and its inverse.

{}Also, instead of metric fluctuations $h_{MN}$, it will be 
convenient to work with ${\widetilde h}_{MN}$ defined via
\begin{equation}
 h_{MN}=\exp(2A) {\widetilde h}_{MN}~.
\end{equation}
It is not difficult to see that in terms of ${\widetilde h}_{MN}$ the
$D$-dimensional diffeomorphisms 
\begin{equation}
 \delta h_{MN}=\nabla_M\xi_N+\nabla_N\xi_M
\end{equation}
are given by the following gauge 
transformations:
\begin{equation}\label{gauge}
 \delta{\widetilde h}_{MN}=\partial_M {\widetilde\xi}_N+
 \partial_N{\widetilde\xi}_M+2A^\prime\eta_{MN}{\widetilde \xi}_S n^S~. 
\end{equation}
Here for notational convenience we have introduced a unit vector $n^M$
with the following components: $n^\mu=0$, $n^D=1$.

\subsection{Equations of Motion} 

{}To proceed further, we need equations of motion for ${\widetilde h}_{MN}$ 
and $\varphi$. Let us
assume that we have matter localized on the brane, and let the corresponding 
conserved energy-momentum tensor be $T_{\mu\nu}$:
\begin{equation}\label{conserved}
 \partial^\mu T_{\mu\nu}=0~.
\end{equation}
The graviton field ${\widetilde h}_{\mu\nu}$ couples to $T_{\mu\nu}$ via
the following term in the action:
\begin{equation}
 S_{\rm {\small int}}=\int_\Sigma d^{D-1} x \left[{1\over 2} T_{\mu\nu}
{\widetilde h}^{\mu\nu}+{8\over{D-2}}\Theta\varphi\right]~,
\end{equation} 
where we have also included the corresponding coupling of $\varphi$ to the
brane matter. Next, starting 
from the action $S+S_{\rm{\small int}}$ we obtain the
following linearized equations of motion for ${\widetilde h_{MN}}$ and
$\varphi$:
\begin{eqnarray}
 &&\left\{\partial_S\partial^S {\widetilde h}_{MN} +\partial_M\partial_N
 {\widetilde h}-\partial_M \partial^S {\widetilde h}_{SN}-
 \partial_N \partial^S {\widetilde h}_{SM}-\eta_{MN}
 \left[\partial_S\partial^S {\widetilde h}-\partial^S\partial^R
 {\widetilde h}_{SR}\right]\right\}+\nonumber\\
 &&(D-2)A^\prime\left\{\left[\partial_S {\widetilde h}_{MN} -
 \partial_M {\widetilde h}_{NS}-\partial_N{\widetilde h}_{MS}\right] n^S
 +\eta_{MN}\left[2\partial^R {\widetilde h}_{RS} - \partial_S 
 {\widetilde h}\right] n^S\right\}-\nonumber\\
 &&\eta_{MN}{\widetilde h}_{SR} n^S n^R V\exp(2A)=\nonumber\\
 &&{8\over {D-2}}\phi^\prime\left[\eta_{MN}\partial_S\varphi n^S-
 \partial_M \varphi n_N-\partial_N\varphi n_M\right]+\eta_{MN}\varphi 
 V_\phi\exp(2A)-M_P^{2-D} {\widetilde T}_{MN}\delta(z)~,\\ 
 &&\partial_S\partial^S\varphi +(D-2) A^\prime\partial_S\varphi n^S-
 {{D-2}\over 8}\varphi V_{\phi\phi}\exp(2A)-{1\over 2}\phi^\prime 
 \left[2\partial^R {\widetilde h}_{RS} - \partial_S 
 {\widetilde h}\right] n^S-\nonumber\\
 &&{{D-2}\over 8} {\widetilde h}_{SR} n^S n^R V_\phi
 \exp(2A)=-M_P^{2-D} {\widetilde \Theta}\delta(z)~,
\end{eqnarray}
where ${\widetilde h}\equiv {\widetilde h}_M^M$, ${\widetilde T}_{MN}\equiv
T_{MN}+T^{\rm{\small brane}}_{MN}$, ${\widetilde \Theta}\equiv\Theta+
\Theta^{\rm{\small brane}}$. Here $T^{\rm{\small brane}}_{MN}$ and 
$\Theta^{\rm{\small brane}}$ are the corresponding brane
contributions (which are linear in ${\widetilde h}_{MN}$ and $\varphi$)
coming from the first term in (\ref{action})\footnote{If the brane 
world-volume theory is not
conformal, then we can {\em a priori} expect that a kinetic term for
the $\phi$ field will also be generated on the brane (just as it happens for  
the graviton). Then $\Theta^{\rm{\small brane}}$ also contains a term
proportional to $\partial^\mu\partial_\mu\varphi$ along with the
term proportional to $f_{\phi\phi}(\phi_0)\varphi$. 
However, at the end of the day we
will find that the brane world-volume theory is conformal, so these
terms are not generated by quantum effects.}. Note that the only
non-vanishing components of ${\widetilde T}_{MN}$ are 
${\widetilde T}_{\mu\nu}$, and we have $\partial^\mu{\widetilde T}_{\mu\nu}=0$.

{}The above equations of motion drastically simplify if we perform a gauge
transformation (\ref{gauge}) with
\begin{equation}\label{special}
 {\widetilde \xi}_M=n_M (\varphi/\phi^\prime)~.
\end{equation}
The new equations of motion then read
\begin{eqnarray}\label{eom1}
 &&\left\{\partial_S\partial^S {\widetilde h}_{MN} +\partial_M\partial_N
 {\widetilde h}-\partial_M \partial^S {\widetilde h}_{SN}-
 \partial_N \partial^S {\widetilde h}_{SM}-\eta_{MN}
 \left[\partial_S\partial^S {\widetilde h}-\partial^S\partial^R
 {\widetilde h}_{SR}\right]\right\}+\nonumber\\
 &&(D-2)A^\prime\left\{\left[\partial_S {\widetilde h}_{MN} -
 \partial_M {\widetilde h}_{NS}-\partial_N{\widetilde h}_{MS}\right] n^S
 +\eta_{MN}\left[2\partial^R {\widetilde h}_{RS} - \partial_S 
 {\widetilde h}\right] n^S\right\}-\nonumber\\
 &&\eta_{MN}{\widetilde h}_{SR} n^S n^R V\exp(2A)=
 -M_P^{2-D} {\widetilde T}_{MN}\delta(z)~,\\ 
 \label{eom2} 
 &&-{1\over 2}\phi^\prime 
 \left[2\partial^R {\widetilde h}_{RS} - \partial_S 
 {\widetilde h}\right] n^S-
 {{D-2}\over 8} {\widetilde h}_{SR} n^S n^R V_\phi
 \exp(2A)=-M_P^{2-D} {\widetilde \Theta}\delta(z)~.
\end{eqnarray}
Note that these equations of motion no longer contain $\varphi$. This has a 
simple physical interpretation \cite{COSM}. 
The domain wall background spontaneously
breaks translational invariance in the $z$ direction. Since this invariance
is a gauge symmetry, the corresponding Goldstone mode, which is given by
configurations where $\omega\equiv\varphi/\phi^\prime$ is independent of $z$
\cite{COSM}\footnote{To see this, note that the translational Goldstone mode
corresponds to fluctuations around the solution given by $\phi(z+\omega(x^\mu))
=\phi(z)+\phi^\prime(z)\omega(x^\mu)+{\cal O}(\omega^2)$.}, must be eaten in 
the corresponding Higgs mechanism. The field which eats the Goldstone mode
is nothing but the graviphoton $h_{\mu D}$ arising in the decomposition of the
$D$-dimensional metric fluctuations in terms of $(D-1)$-dimensional fields
\cite{COSM}.
Note, however, that with the above gauge fixing not only the Goldstone zero
mode but all $\varphi$ modes have been eliminated. There is, however, a price
we have to pay for this simplification. In particular, the residual 
gauge invariance which preserves the equations of motion (\ref{eom1}) and
(\ref{eom2}) is given by
\begin{equation}
 \delta{\widetilde h}_{MN}=\partial_M {\widetilde \xi}_N+\partial_N
 {\widetilde \xi}_M~,~~~
 {\widetilde \xi}_S n^S=0~.
\end{equation}
Note that here ${\widetilde \xi}_M$ need not be independent of $z$. 
Under these residual gauge transformations the fields
${\widetilde h}_{\mu\nu}, A_\mu,\rho$, where 
$A_\mu\equiv{\widetilde h}_{\mu D}$ and $\rho\equiv {\widetilde h}_{DD}$, 
transform as follows
\begin{equation}
 \delta {\widetilde h}_{\mu\nu}=\partial_\mu{\widetilde \xi}_\nu +
 \partial_\nu {\widetilde \xi}_\mu~,~~~
 \delta A_\mu={\widetilde\xi}_\mu^\prime~,~~~\delta\rho=0~.
\end{equation}
This implies that we cannot gauge $\rho$ away. We can, however, gauge
$A_\mu$ away. Thus, in the following we will use the gauge where $A_\mu=0$. 
Note that after this gauge fixing the residual gauge transformations are 
given by
\begin{equation}
 \delta {\widetilde h}_{\mu\nu}=\partial_\mu{\widetilde \xi}_\nu +
 \partial_\nu {\widetilde \xi}_\mu~,~~~\delta\rho=0~,~~~
 {\widetilde\xi}^\prime_\mu=0~.
\end{equation}
We now have the following equations of motion:
\begin{eqnarray}\label{EOM1}
 &&\left\{\partial_\sigma\partial^\sigma 
 H_{\mu\nu} +\partial_\mu\partial_\nu
 H-\partial_\mu \partial^\sigma H_{\sigma\nu}-
 \partial_\nu \partial^\sigma H_{\sigma\mu}-\eta_{\mu\nu}
 \left[\partial_\sigma\partial^\sigma H-\partial^\sigma\partial^\rho
 H_{\sigma\rho}\right]\right\}+\nonumber\\
 &&\left\{H_{\mu\nu}^{\prime\prime}-\eta_{\mu\nu}H^{\prime\prime}+
 (D-2)A^\prime\left[H_{\mu\nu}^\prime-\eta_{\mu\nu}H^\prime\right]\right\}+
 \nonumber\\
 &&\left\{\partial_\mu\partial_\nu\rho-\eta_{\mu\nu}
 \partial_\sigma\partial^\sigma 
 \rho+\eta_{\mu\nu}\left[(D-2)A^\prime\rho^\prime
 -\rho V\exp(2A)\right]\right\}= -M_P^{2-D} {\widetilde T}_{\mu\nu}
 \delta(z)~,\\ 
 \label{EOM2} 
 &&\left[\partial^\mu H_{\mu\nu}-\partial_\nu H\right]^\prime +(D-2)A^\prime 
 \partial_\nu\rho=0~,\\
 \label{EOM3}
 &&-\left[\partial^\mu\partial^\nu H_{\mu\nu}-\partial^\mu\partial_\mu H\right]
 +(D-2) A^\prime H^\prime +\rho V\exp(2A)=0~,\\
 \label{EOM4}
 &&\phi^\prime \left[H^\prime-\rho^\prime\right]-
 {{D-2}\over 4} \rho V_\phi\exp(2A)=
 -2 M_P^{2-D}{\widetilde\Theta}\delta(z)~,
\end{eqnarray}
where $H_{\mu\nu}\equiv {\widetilde h}_{\mu\nu}$, and $H\equiv H_\mu^\mu$.

{}Not all of the above equations are independent. First, differentiating
(\ref{EOM1}) with $\partial^\mu$ (and taking into account that $\partial^\mu
{\widetilde T}_{\mu\nu}=0$), we obtain an equation which is identically 
satisfied once we take into account (\ref{EOM2}) together with the
equations of motion for $A$ and $\phi$. Next, taking the trace in (\ref{EOM1}),
we obtain an equation which together with (\ref{EOM2}), ({\ref{EOM3}) and
the equations of motion for $A$ and $\phi$ gives the following equation:
\begin{equation}
 {4\over{D-2}}(\phi^\prime)^2\left[H^\prime-\rho^\prime\right]-\rho V_\phi
 \phi^\prime\exp(2A)
 =M_P^{2-D} A^\prime {\widetilde T}\delta(z)~,
\end{equation} 
where ${\widetilde T}\equiv {\widetilde T}_\mu^\mu$. This equation is 
compatible with (\ref{EOM4}) if and only if 
\begin{equation}\label{condition0}
 {\widetilde\Theta}=-{{D-2}\over 8}{A^\prime(0)\over\phi^\prime(0)}
 {\widetilde T}~.
\end{equation}
Thus, we already see that the coupling of the brane matter to the bulk scalar
cannot be arbitrary but is determined in terms of the trace of the 
energy-momentum tensor. (Note that neither $A^\prime$ nor $\phi^\prime$ vanish
anywhere in the backgrounds we consider here, including the location of the 
brane $z=0$.)

{}Finally, let us discuss (\ref{EOM4}). It came from the equation of motion
for $\varphi$, which was a second order equation. However, after we eliminated
$\varphi$ itself, this equation became a first order equation in terms of
$H$ and $\rho$. This then implies that the source term on the r.h.s. of
(\ref{EOM4}) must vanish or else $H-\rho$ 
will be discontinuous. Thus, we have
arrived at the conclusion that consistency of the above equations implies 
that we must have
\begin{equation}\label{condition}
 {\widetilde T}={\widetilde \Theta}=0~.
\end{equation}
This, in particular, implies that the brane world-volume theory is
generically expected to be conformal 
in this setup. Indeed, it is not difficult to see
that (\ref{condition}) can be satisfied if   
$T\equiv T^\mu_\mu$ as well as $\Theta$ vanish.  
To ensure conformality of the matter localized on the brane
then generically requires that the brane world-volume theory itself
be conformal. On the other hand, if this is not the case then 
to satisfy (\ref{condition})
$T$ and $\Theta$ (in the best case where $f_{\phi\phi}(\phi_0)=0$)
must be the same up to a non-vanishing constant\footnote{To see this, note 
that, once we perform the gauge transformation (\ref{gauge}) with 
the gauge parameter given in (\ref{special}), ${\widetilde T}_{\mu\nu}$
contains a term proportional to $(D-3)\left[\partial_\mu\partial_\nu\omega -
\eta_{\mu\nu}\partial^\sigma\partial_\sigma\omega\right]$. Assuming 
$D\not=2,3$, we then have that on the brane
$\partial^\sigma\partial_\sigma\omega$ is
proportional to $T$ with a non-vanishing coefficient. On the other
hand, the condition ${\widetilde \Theta}=0$ gives a second order differential
equation on the brane for $\omega$ with a source term
proportional to $\Theta$. Hence the aforementioned conclusion about the
relation between $T$ and $\Theta$.}, 
which generically need not be the case.   

{}Let us verify that, if (\ref{condition}) is satisfied, the above
system of equations does have a 
consistent solution for $H_{\mu\nu}$ and $\rho$. It is not difficult to 
show that such a solution indeed exists, and is given by (note
that $p^\mu{\widetilde T}_{\mu\nu}(p)=0$)
\begin{eqnarray}
 &&\rho=0~,\\
 &&H_{\mu\nu} (p,z) =M_P^{2-D} {\widetilde T}_{\mu\nu}(p)\Omega(p,z)~,
\end{eqnarray} 
where we have performed a Fourier transform w.r.t. the coordinates $x^\mu$
(and the corresponding momenta are $p^\mu$). Also, let us Wick rotate to the
Euclidean space (where the propagator is unique).
The function $\Omega(p,z)$ is
a solution to the following equation ($p^2\equiv p^\mu p_\mu$)
\begin{equation}
 \Omega^{\prime\prime}(p,z)+(D-2)A^\prime\Omega^\prime(p,z)-p^2\Omega(p,z)=-
 \delta(z)
\end{equation}
subject to the boundary conditions (for $p^2>0$)
\begin{equation}
 \Omega(p,z\rightarrow\pm \infty)=0~.
\end{equation}
The above solution describes a gravitational field of conformal matter
localized on the brane.

\subsection{Additional Evidence}

{}In this subsection we would like to give additional evidence that
the condition ${\widetilde\Theta}=0$ (from which it follows that 
${\widetilde T}=0$) is indeed necessary. 
To begin with, let us perform the aforementioned Fourier transform
in (\ref{EOM1}), (\ref{EOM2}), (\ref{EOM3}) and (\ref{EOM4}), and Wick rotate
to the Euclidean space. The equations of motion for the Fourier transformed
quantities read:
\begin{eqnarray}\label{EOM1F}
 &&-\left\{p^2 
 H_{\mu\nu} +p_\mu p_\nu
 H-p_\mu p^\sigma H_{\sigma\nu}-
 p_\nu p^\sigma H_{\sigma\mu}-\eta_{\mu\nu}
 \left[p^2 H-p^\sigma p^\rho
 H_{\sigma\rho}\right]\right\}+\nonumber\\
 &&\left\{H_{\mu\nu}^{\prime\prime}-\eta_{\mu\nu}H^{\prime\prime}+
 (D-2)A^\prime\left[H_{\mu\nu}^\prime-\eta_{\mu\nu}H^\prime\right]\right\}+
 \nonumber\\
 &&\left\{-p_\mu p_\nu\rho
 +\eta_{\mu\nu}\left[(D-2)A^\prime\rho^\prime +\rho \left(p^2
 -V\exp(2A)\right)\right]\right\}= -M_P^{2-D} {\widetilde T}_{\mu\nu}(p)
 \delta(z)~,\\ 
 \label{EOM2F} 
 &&\left[p^\mu H_{\mu\nu}-p_\nu H\right]^\prime +(D-2)A^\prime 
 p_\nu\rho=0~,\\
 \label{EOM3F}
 &&\left[p^\mu p^\nu H_{\mu\nu}-p^2 H\right]
 +(D-2) A^\prime H^\prime +\rho V\exp(2A)=0~,\\
 \label{EOM4F}
 &&{4\over{D-2}}(\phi^\prime)^2 \left[H^\prime-\rho^\prime\right]-
 {{D-2}\over 4} \rho V_\phi\phi^\prime \exp(2A)=
 M_P^{2-D}A^\prime {\widetilde T}(p)\delta(z)~,
\end{eqnarray}
where we have taken into account (\ref{condition0}). 

{}Let us assume that ${\widetilde T}(p)\not=0$.
Then the most general tensor structure for the fields $H_{\mu\nu}$ and
$\rho$ can be parametrized in terms of four functions $a,b,c,d$ as follows:
\begin{eqnarray}
 &&\rho=M_P^{2-D}~d ~{\widetilde T}(p)~,\\
 &&H_{\mu\nu}=M_P^{2-D}
 \left\{a~{\widetilde T}_{\mu\nu}(p) +\left[b~\eta_{\mu\nu}+c~p_\mu p_\nu
 \right]{\widetilde T}(p)\right\}~.
\end{eqnarray}
Plugging this back into the equations of motion, we obtain six equations
for four unknowns $a,b,c,d$. However, as should be clear from the above 
discussion, two of them are identically satisfied once we take into account 
the other four (as well as the equations of motion for $A$ and $\phi$). 
After some straightforward computations we obtain the following
system of four independent equations:
\begin{eqnarray}
 &&a^{\prime\prime}+(D-2)A^\prime a^\prime-p^2 a=-\delta(z)~,\\
 &&(D-2)A^\prime d=a^\prime+(D-2)b^\prime~,\\
 &&A^\prime\left[(D-2)p^2c^\prime-a^\prime\right]=p^2
 \left[a+(D-2)b\right]-{4\over{D-2}}(\phi^\prime)^2 d~,\\
 &&a+(D-3)b-c^{\prime\prime}-(D-2)A^\prime c^\prime+d=0~.
\end{eqnarray}
Note that from the first equation it follows that $a(p,z)=\Omega(p,z)$.

{}Let $w\equiv a+(D-2)b$. Using the above equations for $a,b,c,d$ after some
straightforward computations we obtain the following second order equation for
$w$:
\begin{equation}\label{w}
 w^{\prime\prime}+\left\{(D-2)A^\prime-\left[\ln(F)\right]^\prime\right\}
 w^\prime-p^2 w=-F\delta(z)~,
\end{equation} 
where $F(z)$ is the following function:
\begin{equation}
 F\equiv{{(A^\prime)^2}\over{(A^\prime)^2-A^{\prime\prime}}}~.
\end{equation}
Solutions to the above equation for $w$ have some peculiar properties.
To expose them, we need to study the asymptotic behavior of the function
$\left[\ln(F)\right]^\prime$.

{}To begin with, it is not difficult to see that 
\begin{equation}
 F=-{\beta\over\alpha}{W^2\over (W_\phi)^2}~,
\end{equation}
where the superpotential $W$ as well as constants $\alpha,\beta$ were defined
in section II. It then follows that
\begin{equation}
 \left[\ln(F)\right]^\prime=2\alpha\exp(A){{(W_\phi)^2-WW_{\phi\phi}}\over W}~.
\end{equation}
Let us compute this function in the example discussed at the end of section II.
In that example the superpotential is given by (\ref{sup1}). We then have
\begin{equation}
 \left[\ln(F)\right]^\prime=-2\alpha\xi v\exp(A)
 {{3+2{\widetilde \phi}+{\widetilde
 \phi}^2}\over{2+{\widetilde \phi}}}~,
\end{equation}
where ${\widetilde\phi}\equiv\phi/v$. For definiteness let us assume that
$\alpha\xi v>0$. Then at $z\rightarrow+\infty$ the domain wall solution 
approaches the Minkowski vacuum where ${\widetilde\phi}=+1$, while at
$z\rightarrow -\infty$ 
it approaches the AdS vacuum where ${\widetilde\phi}=-1$.
It then follows that $\left[\ln(F)\right]^\prime$ is always negative
on the solution. Moreover, at $z\rightarrow+\infty$ (where $A$ goes to a
constant and $A^\prime$ goes to zero) $\left[\ln(F)\right]^\prime$ goes
to a constant, which we will denote by $-2\zeta$. Then for large positive $z$
$w$ is well approximated by the solution to the following equation
\begin{equation}
 w^{\prime\prime}+2\zeta w^\prime-p^2 w=0
\end{equation}
subject to the boundary condition $w(z\rightarrow+\infty)=0$. This solution
is given by
\begin{equation}
 w(z)={\rm const.}\times\exp(-\lambda z)~,
\end{equation}
where (the other root of the corresponding quadratic equation is negative)
\begin{equation}
 \lambda\equiv \zeta +\sqrt{\zeta^2+p^2}~.
\end{equation}
Note that for $p\rightarrow 0$ $\lambda$ does not vanish but approaches 
$2\zeta$. This implies that even at zero momentum there is a 
non-trivial solution to
(\ref{w})\footnote{Here we should point out that this does not occur
if we consider domain walls arising for runaway type of potentials
discussed in \cite{zura1}. Nonetheless, the fact that a non-trivial
solution of (\ref{w}) does exist in the above example if we assume 
${\widetilde T}\not=0$ might be considered as (at least indirect) evidence
that ${\widetilde T}=0$ condition is indeed necessary. At any rate,
if this condition is not satisfied, as we have already pointed out, there is
a discontinuity in $H-\rho$, which appears to be inconsistent.}. 
In particular, it is given by $w(z)={\widetilde w}(0)$ for $z<0$,
$w(z)={\widetilde w}(z)$ for $z\geq 0$, where ${\widetilde w}(z)$ is the 
solution of the equation
\begin{equation}
 {\widetilde w}^{\prime\prime}+
 \left\{(D-2)A^\prime-\left[\ln(F)\right]^\prime\right\}
 {\widetilde w}^\prime-p^2 {\widetilde w}=0
\end{equation}
subject to the boundary conditions ${\widetilde w}(z\rightarrow+\infty)=0$,
and ${\widetilde w}^\prime (0)=-F(0)$ (note that $F(z)$ is always positive). 
The fact that such a solution always 
exists for $v\ll 1$ can be seen from the fact that in this case 
$(D-2)A^\prime \ll -\left[\ln(F)\right]^\prime$.

{}The existence of a non-trivial solution at $p^2=0$ indicates an inconsistency
in the system. 
Note that if we have ${\widetilde T}=0$ to begin with, 
then we do not have the same system of equations for $a,b,c,d$ as above. 
In fact in this case there is no inconsistency, and we have a consistent 
solution discussed in the previous subsection.

\section{Remarks}

{}Thus, as we see, in the setup of \cite{zura1}, where the brane cosmological 
constant is protected by bulk supersymmetry, to have consistent couplings
of the bulk scalar and gravity to matter localized on the brane, it appears to
be necessary that the latter is conformal. This then implies that (generically)
the brane world-volume theory should itself be
conformal. The fact that the brane 
cosmological constant vanishes is then a trivial consequence of conformal 
invariance of the brane world-volume theory. However, what appears to be 
non-trivial is that unbroken bulk supersymmetry in this setup (where the 
volume of the extra dimension is infinite) actually protects conformality of
the brane world-volume theory (which {\em a priori} need not even be 
supersymmetric).

{}In this context one might hope to use this setup as a possible realization of
the conformal approach to phenomenology \cite{FV} (also see \cite{BKV}).
However, it is still unclear how one could possibly have the brane
conformal invariance broken around TeV while having much larger Planck scale 
on the brane.

\acknowledgments

{}I would like to thank Gregory Gabadadze, Juan Maldacena, Tom Taylor
and Cumrun Vafa for
valuable discussions. Parts of this work were completed while I was
visiting at Harvard University and New York University.
This work was supported in part by the National Science Foundation.
I would also like to thank Albert and Ribena Yu for financial support.

\end{document}